\documentclass[mathleft,fleqn,%
]{an}
\usepackage{graphicx}
\usepackage[varg]{txfonts}
\newcommand{\msun}{$M_\odot$}

\begin{document}

\Pagespan{1}{}
\Yearpublication{2017}%
\Yearsubmission{2016}%
\Month{1}%
\Volume{338}%
\Issue{1}%
\DOI{asna.201700000}%

\title{The cataclysmic variable AE Aquarii:  B-V colour of the  flares
\thanks{Observations from National Astronomical Observatory  Rozhen and AO Belogradchick, Bulgaria}}

\author{R.\,K.\,Zamanov\inst{1}\fnmsep\thanks{Corresponding author:
        {rkz@astro.bas.bg,   glatev@astro.bas.bg }}
\and  G.\,Y. Latev\inst{1}
\and  S. Boeva\inst{1}
\and  S. Ibryamov\inst{2}
\and  G. B. Nikolov\inst{1}
\and  K. A. Stoyanov\inst{1}
}
\titlerunning{Flares of AE Aqr}
\authorrunning{Zamanov et al.}
\institute{
Institute of Astronomy and National Astronomical Observatory, Bulgarian Academy of Sciences, 
           Tsarigradsko shose 72, BG-1784, Sofia, Bulgaria 
\and 
Department of Physics, Konstantin Preslavsky University of Shumen, 115 Universitetska Str., 9700 Shumen, Bulgaria
}

\received{1 September 2016}
\accepted{19 December 2016}
\publonline{.. ... 2017}


\abstract{   We report simultaneous observations of the flaring behaviour of the cataclysmic variable star AE~Aqr.
The observations are in Johnson B and V bands. The colour-magnitude diagrams 
(B-V versus V  and  B-V vs. B) show that the star becomes blues as it becomes brighter. 
In our model AE~Aqr behaviour can be explained with  flares (fireballs) 
with $0.03 \le B-V \le 0.30$ and temperature in the interval 
$8000 \le T \le 12000$.  }

\maketitle

\section{Introduction}
AE Aqr is  a bright ($V \sim 11$ mag) unusual cataclysmic variable, displaying strong flaring activity
(e.g. Chincarini \& Walker 1974). Its  binary nature was discovered by Joy (1954). 
In  the system a spotted K type dwarf  (K0-K4 IV/V star) 
transfers material through the inner Lagrangian point L$_1$ 
toward  a magnetic white dwarf  (Skidmore et al. 2003,  Hill et al. 2016).
It has a relatively long (for a cataclysmic variable) orbital period 
of 9.88~h (e.g. Casares et al. 1996) and a very short spin period of the white dwarf of only 33~s,  
detected  in the optical and X-ray bands (Patterson et al. 1980). 
To appear in such a state, AE~Aqr should be a former supersoft X-ray binary, in which the mass transfer rate 
in the recent past ($\approx 10^7$~yr) has been  much higher than its current value (Schenker et al. 2002).  

The high-dispersion time-resolved absorption line spectroscopy by  Echevarr{\'{\i}}a et al. (2008)
gives the binary parameters as:  
white dwarf mass  $M_{WD} = 0.63 \pm 0.05$ \msun,  
secondary mass $M_2 = 0.37 \pm 0.04 $ \msun, 
binary separation $a = 2.33 \pm 0.02$ R$_\odot$, 
and inclination $i \approx 70^o$.


The observations obtained till now  contain no clear evidence for an accretion disc.  
Given the lack of  such evidence,    
speedy magnetic propeller (Eracleous \& Horne 1996; Wynn, King \& Horne 1997) and 
ejector  (Ikhsanov, Neustroev \& Beskrovnaya 2004; Ikhsanov \& Beskrovnaya  2012) 
are supposed to operate in this star. 
The hydrodynamical calculations hint that the field does not hinder the formation
of a transient disk (ring) surrounding the magnetosphere (Isakova et al. 2016).
AE~Aqr also exhibits radio and millimeter synchrotron emission 
(e.g. Bookbinder \& Lamb 1987; Bastian, Dulk \& Chanmugam~1988).
XMM-Newton observations demonstrate that  the plasma cannot be a product of mass accretion onto the white dwarf
(Itoh et al. 2006).
Non-detection of very high energy  (TeV) $\gamma$-rays in MAGIC observations 
(Aleksi{\'c} et al. 2014)  probably points that the white dwarf is not acting as ejector 
and from our point of view the propeller model is the better one.

In the optical bands AE Aqr exhibits  atypical flickering consisting of  
large optical flares with amplitude $\le 1$ magnitude in Johnson V band.
The flares are visible about one third of the time and have 
typical rise time $200 - 400$~s (van Paradijs, van Amerongen  \& Kraakman, 1989; 
Zamanov et al. 2012). 
The flares are also visible in the X-rays (Mauche et al. 2012) and 
in submillimeter wavelengths (Torkelsson 2013).

Here we explore the behaviour of AE~Aqr in optical B and V bands, 
construct the colour-magnitude diagram, 
and estimate the  B-V colour of the flares and temperature of the fireballs.

\section{Observations}

\begin{figure*}    
   \vspace{9.8cm}     
   \includegraphics{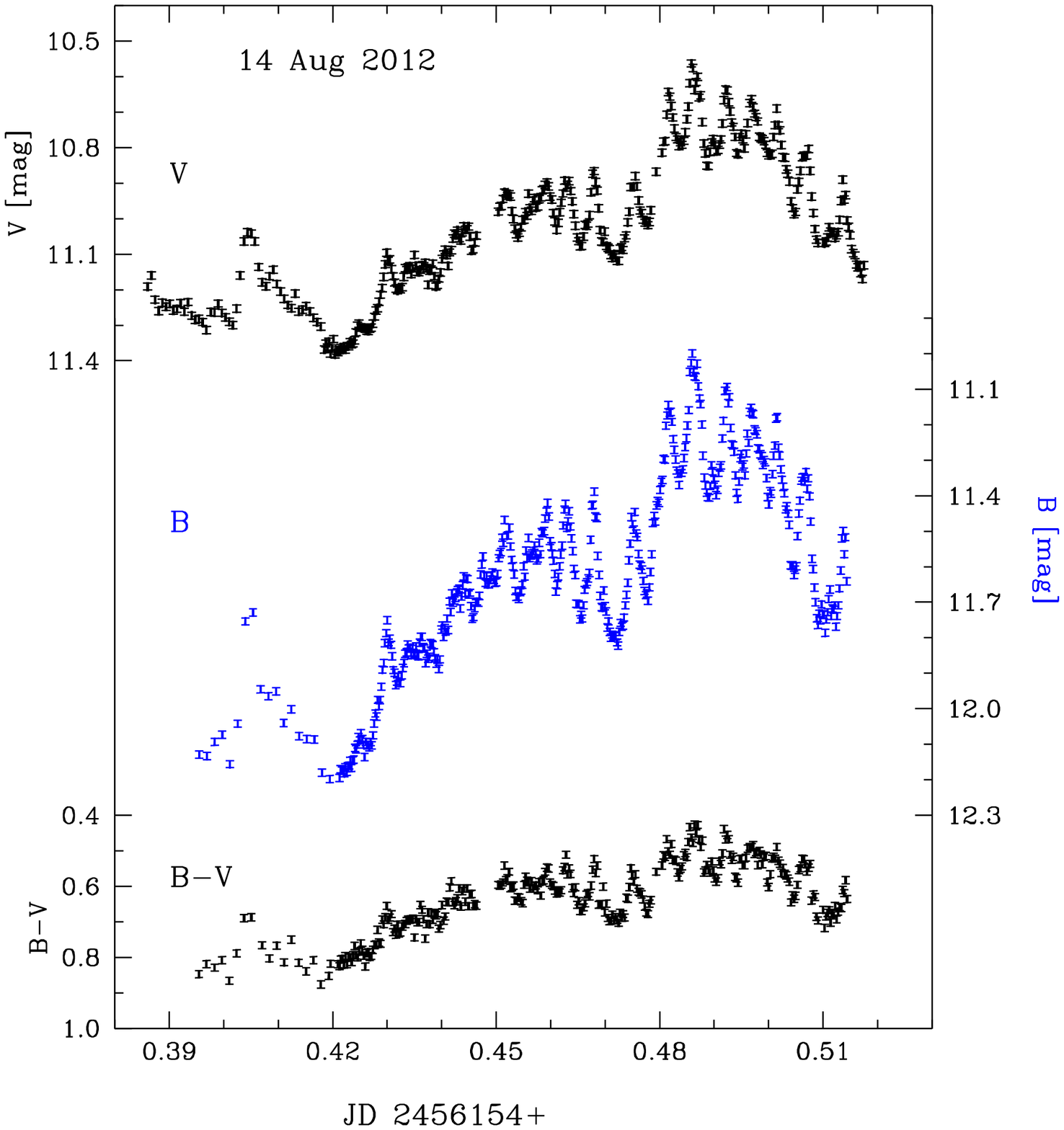}      
   \includegraphics{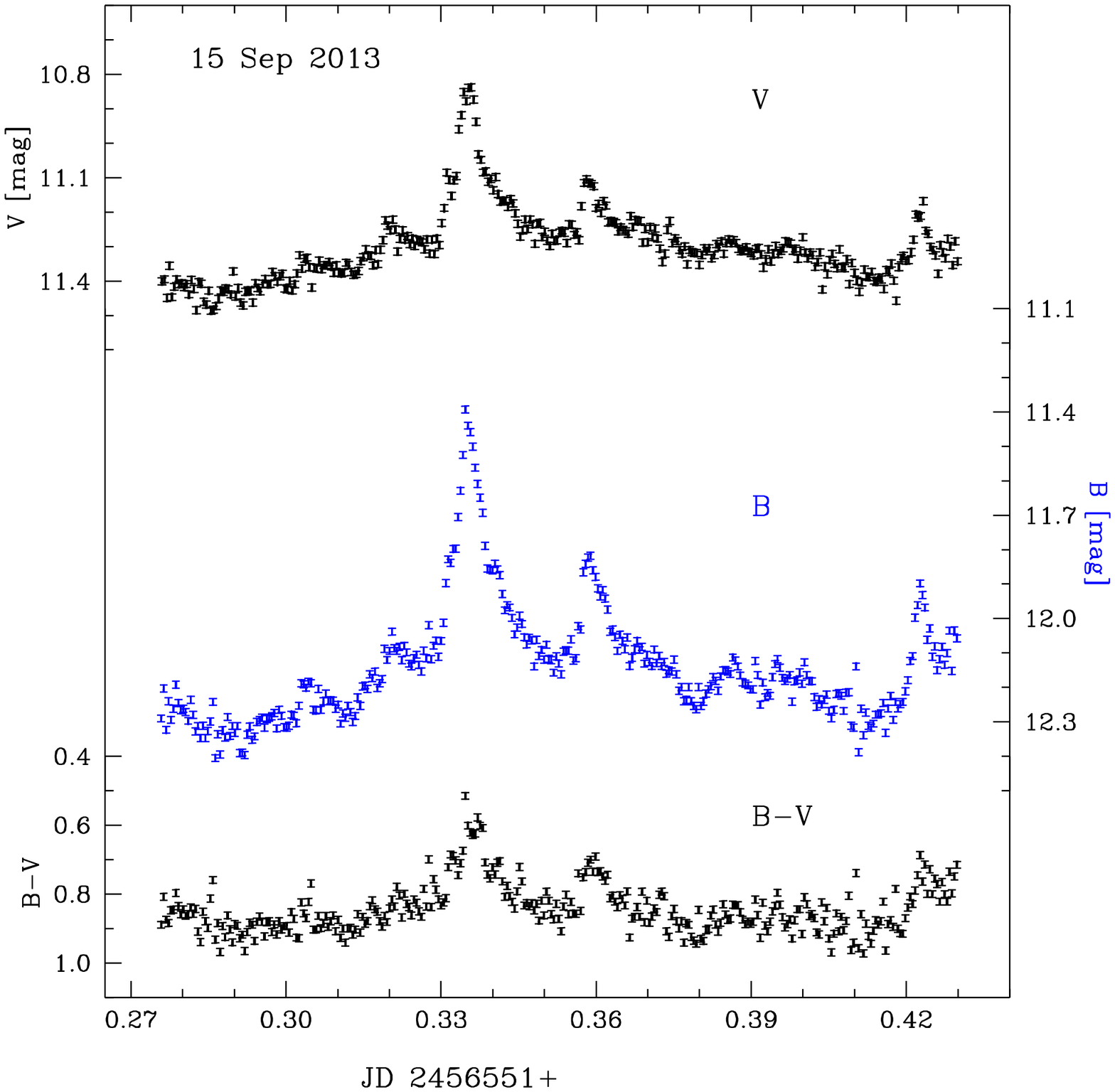}      
   \caption[]{Simultaneous observations of AE~Aqr in B and V bands obtained on August 14, 2012
   and September 15, 2013.  The calculated B-V colour index is also plotted.
   When the brightness of the star increases B-V index becomes bluer. 
   }
   \label{fig1}    
 \end{figure*}

The observations of the short term variability of AE~Aqr 
are obtained with 4 telescopes: 
the 2.0 m RCC telescope, 
the 50/70 cm Schmidt telescope,
the 60 cm  telescope of the Bulgarian National Astronomical Observatory  Rozhen, 
and the 60 cm telescope of the Belogradchick Astronomical Observatory. 
All of the telescopes are equipped with CCD cameras. 
All the CCD images have been bias subtracted, flat fielded, and standard 
aperture photometry has been performed. The data reduction and aperture photometry 
were done  with \textsc{iraf} and checked with alternative software packages. 
The typical accuracy of the photometry was 0.005-0.010 mag. 
Journal of observations is given in Table~1.  
The first four dates are partly analyzed in Zamanov et al. (2012). 
The other are new observations. Two examples of our observations are presented in Fig.~\ref{fig1}. 


\begin{table*}
  \begin{center}
  \caption{Simultaneous observations of AE~Aqr in B and V bands. 
  The columns give  date of observation (in format YYYYMMDD),
  duration of the run (in minutes), 
  telescope, exposure times of the CCD frames, number of the data points in the run for B and V bands. 
  The last column gives the number of the data points over which the B-V colour is calculated. 
   }
  \begin{tabular}{ l l | lrr | llrr | crllcccrrr}
  date     & D   &	\multicolumn{3}{|c|}{B} 	    & & \multicolumn{3}{c|}{V}  	  &           &  \\
           & min &  telescope  & exp-time	 & $N_{B}$  & &  telescope   & exp-time & $N_{V}$ & $N_{B-V}$ &  \\
  \hline
           &     &             &                 &          & &           &           &         &          &   \\
  20100813 & 162 &   50/70Sch  &  60s		 &   137    & & 2.0m	  &   5s,10s  &   463	& 110	   &   \\
  20100814 & 299 &   50/70Sch  &  60s,120s	 &   246    & & 2.0m	  &   10s     &  1013	& 228	   &   \\
  20100816 & 353 &   60Roz     &  60s,90s	 &   75     & & 60Roz	  &   60s     &    68	&  60	   &   \\
  20110831 & 180 &   50/70Sch  &  30s		 &   270    & & 2.0m	  &    5s     &  1087	& 258	   &   \\
  20110924 & 166 &   60Bel     &  20s		 &   160    & & 60Bel	  &   10s     &   160	& 160	   &   \\
  20110927 & 164 &   60Bel     &  15s		 &   200    & & 60Bel	  &   10s     &   200	& 200	   &   \\
  20110928 & 123 &   60Bel     &  15s		 &   150    & & 60Bel	  &   10s     &   150	& 150	   &   \\
  20120814 & 189 &   50/70Sch  &  15s		 &   442    & & 2.0m,60R  &   5s,10s  &   395	& 327      &   \\
  20120815 & 167 &   60Roz     &  30s		 &   158    & & 50/70Sch  &   10s     &    75	&  60	   &   \\
  20120816 & 121 &   50/70Sch  &  15s		 &   380    & & 2.0m	  &   3s,10s  &   282	& 282      &   \\
  20120908 & 107 &   50/70Sch  &  20s		 &   190    & & 60Roz	  &   10s     &   347	& 167	   &   \\
  20130915 & 221 &   50/70Sch  &  10s		 &   325    & & 50/70Sch  &    5s     &   325	& 325	   &   \\
  \\
  \end{tabular}
  \label{tab1}
  \end{center}
\end{table*} 

\section{Light curves in B and V bands}

The stochastic light variations on timescales of a few minutes (flickering) with amplitude of a few$\times0.1$ magnitudes 
is a type  of variability observed in the three main classes of binaries
that contain white dwarfs  accreting  material from a companion mass-donor star:  
cataclysmic variables (CVs), supersoft X-ray binaries, 
and symbiotic stars (e.g. Sokoloski 2003). 
Stochastic light variations are not only observed in accreting white dwarfs, but also in  
accreting black holes and neutron stars (e.g. Belloni \& Stella 2014; Scaringi 2015).  


The flickering of  AE~Aqr is  first detected by Henize (1949). 
Five-colour optical photometry in Walraven system made in 1984 and 1985 was analyzed by van Paradijs et al. (1989).  
In Fig.~\ref{fig1} is plotted  the variability of AE Aqr in B and V bands. The calculated
colour index is also plotted. 
All data are drawn on identical scale, in this way 0.1 magnitude  
has the same size on all Y axes. 
  
 \begin{figure*}    
   \vspace{9.0cm}     
   \includegraphics{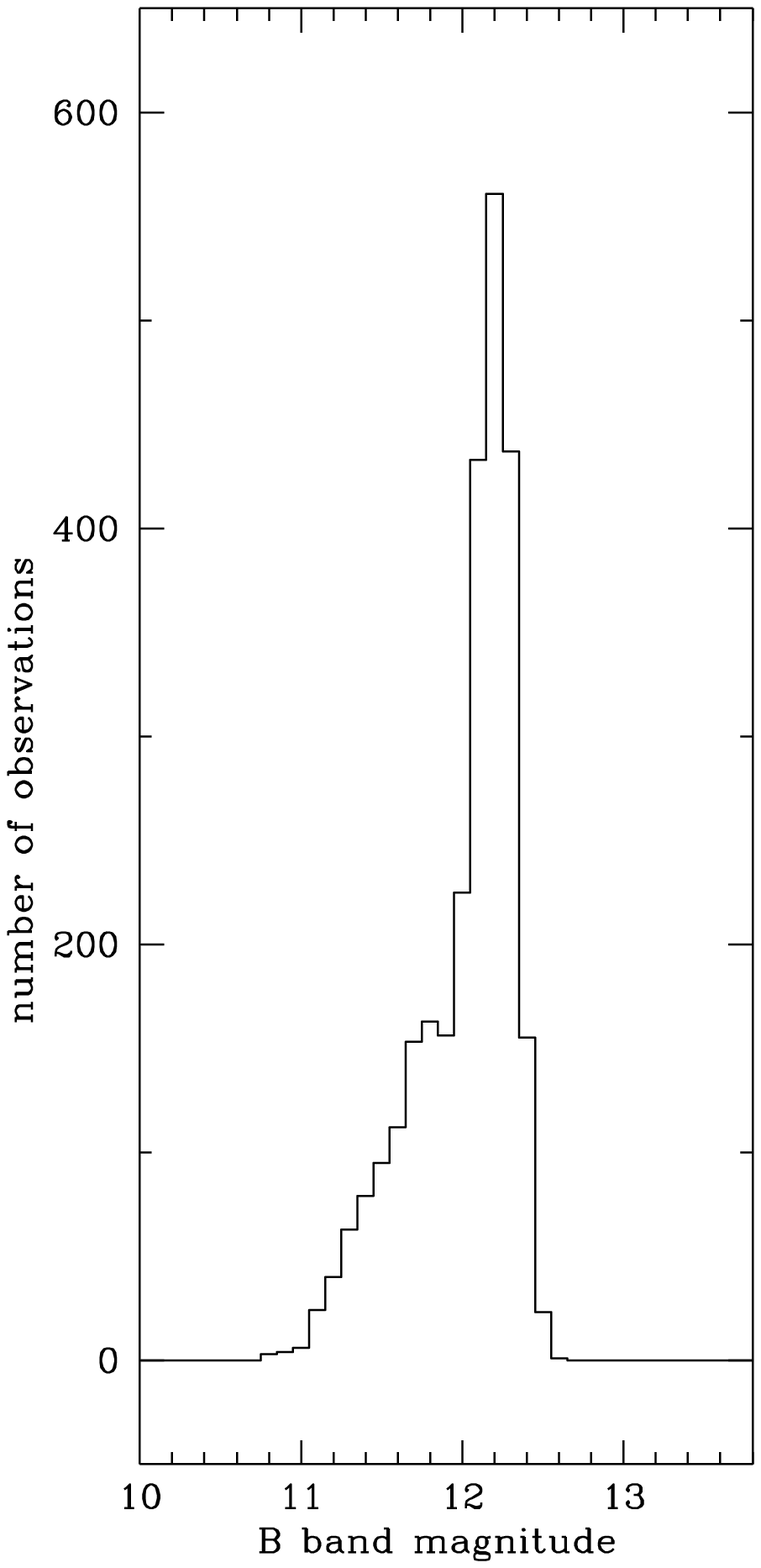}      
   \includegraphics{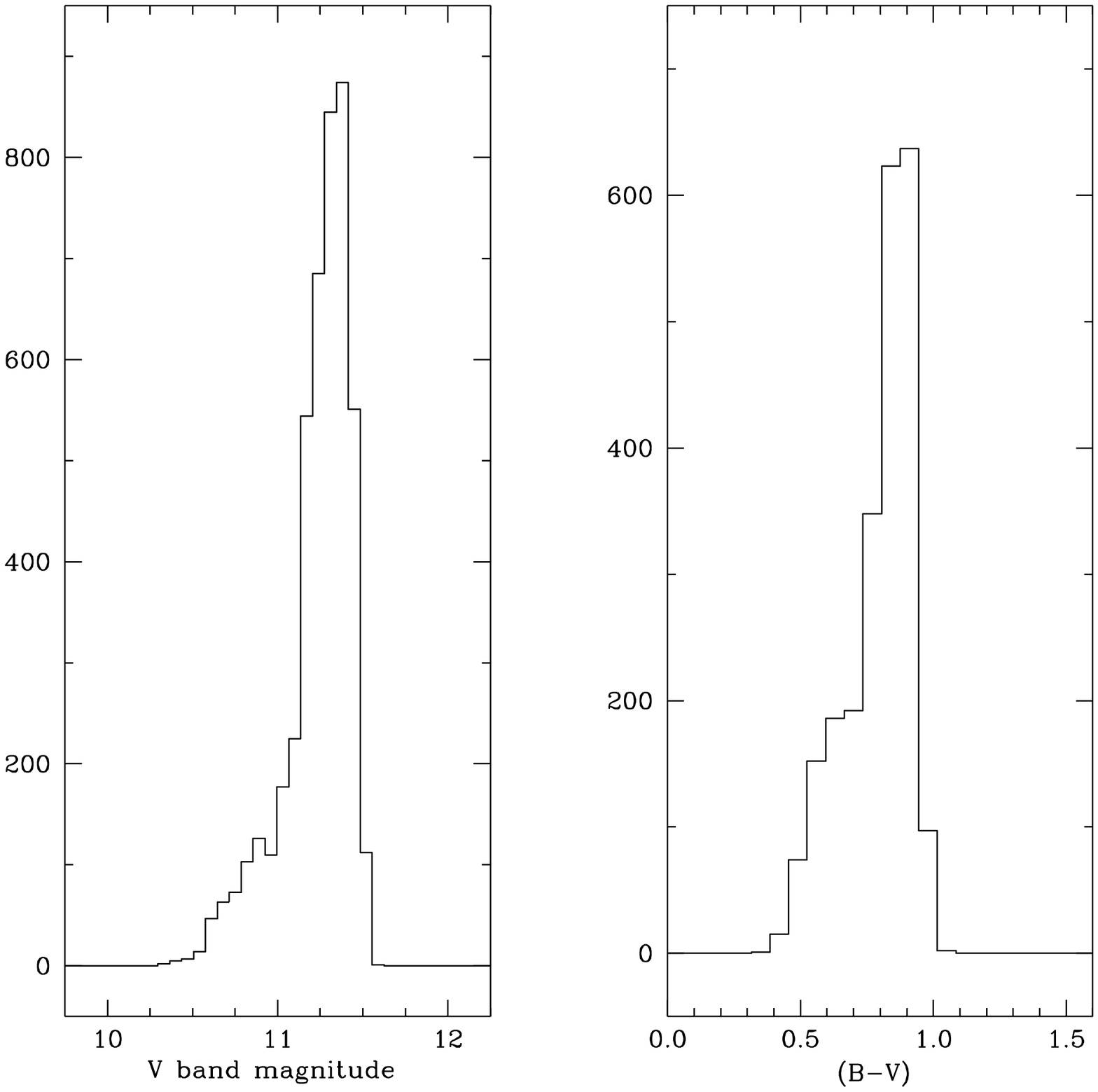}      
   \caption[]{Histograms representing the distribution of B and V band magnitudes and $B-V$ colour.
    The total number is 2473  data points for which  V, B  and $B-V$ are calculated. }
   \label{fig.hist}    
 \end{figure*}	     

Histograms representing the distribution of B and V band magnitudes and $B-V$ colour
are plotted in Fig.~\ref{fig.hist}. 
B magnitude is measured for  $N_{pts}=2733$. The values are  in the interval  $10.767 \le B \le 12.565$, with average value 
$\overline{B}  = 12.006$,  standard deviation of the mean $\sigma_B=0.320$ and median value  $<B> = 12.112$. 
V magnitude is measured for  $N_{pts}= 4564$. The values are the interval  $10.355 \le V \le 11.563$, with average value 
$\overline{V}  = 11.235$,  $\sigma_V= 0.199$  and median value  $<V> \: = \: 11.288$. 
The histograms  of B and V  band magnitudes  have a well defined peak and extended tail to the higher brightness. 
The peaks of the distributions are at $V=11.35 \pm 0.03$ and at $B=12.22 \pm 0.03$ for the V and B band respectively.   
$B-V$ is measured for  $N_{pts}= 2327$ data points. The values are the interval 
$0.354 \le B-V \le 1.037$, with average value 
$\overline{B-V}  =  0.794$,  $\sigma_{B-V} = 0.126$ and median value  $<B-V>=0.829$.

The histogram  of $B-V$  colour have a well defined peak at $B-V=0.88 \pm 0.03$ and extended tail to the blue 
colour. The tails in all three histograms plotted in 
Fig.~\ref{fig.hist} are due to the appearance of the flares in the light curve.

\section{Colour-magnitude diagram }

\begin{figure*}    
   \vspace{19.0cm}     
    \includegraphics{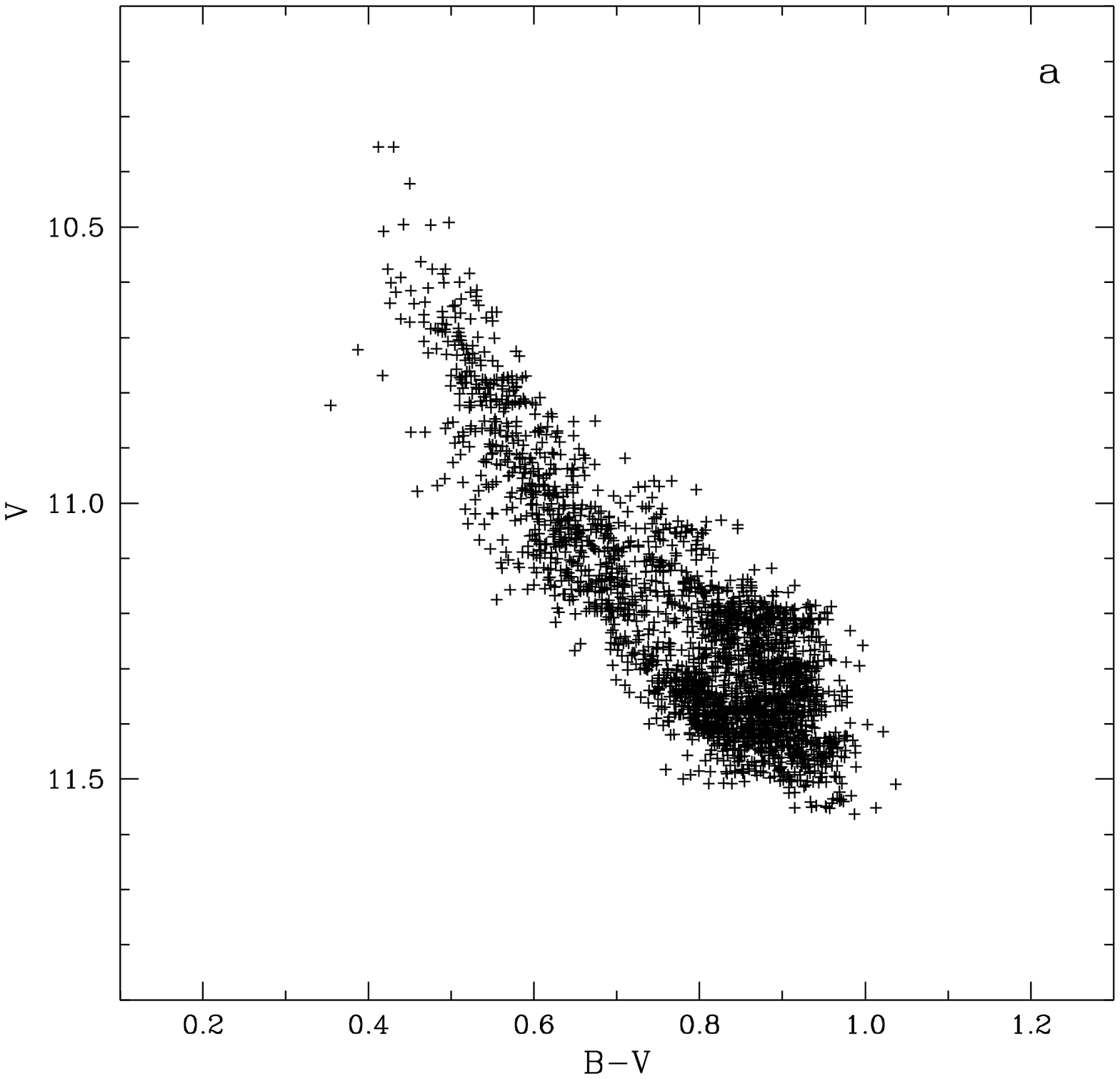}	  
    \includegraphics{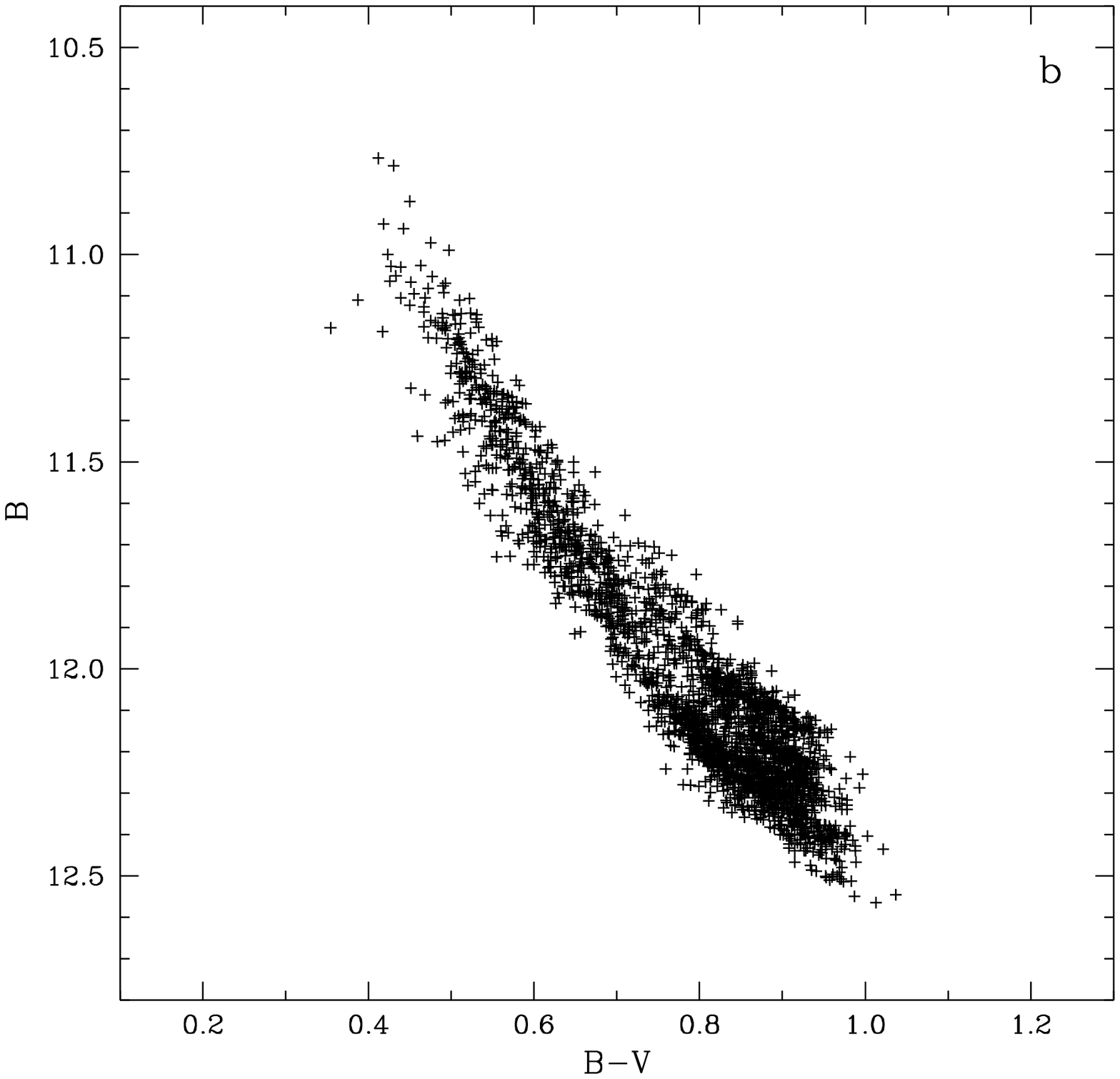}	  
    \includegraphics{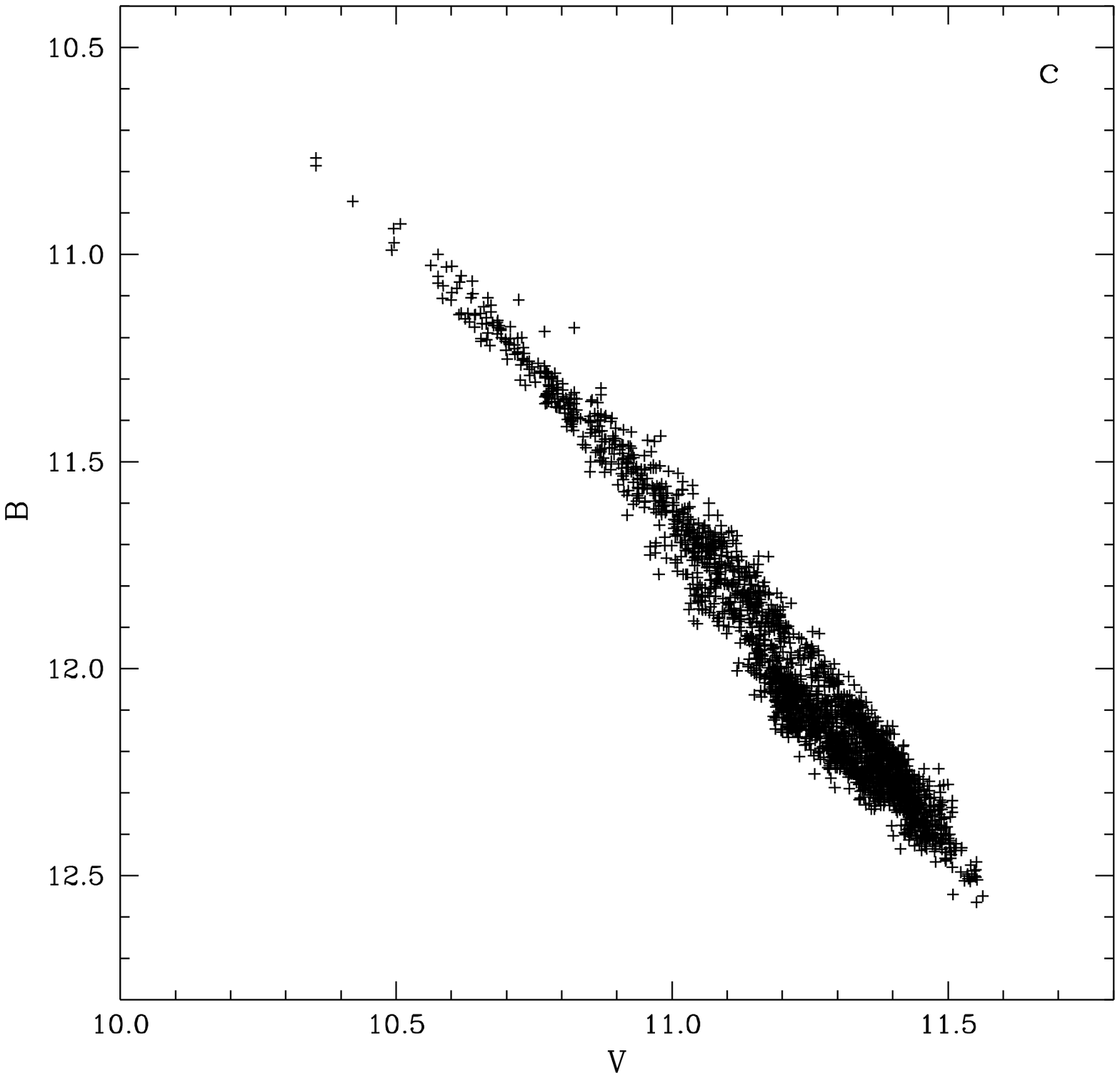}	  
    \includegraphics{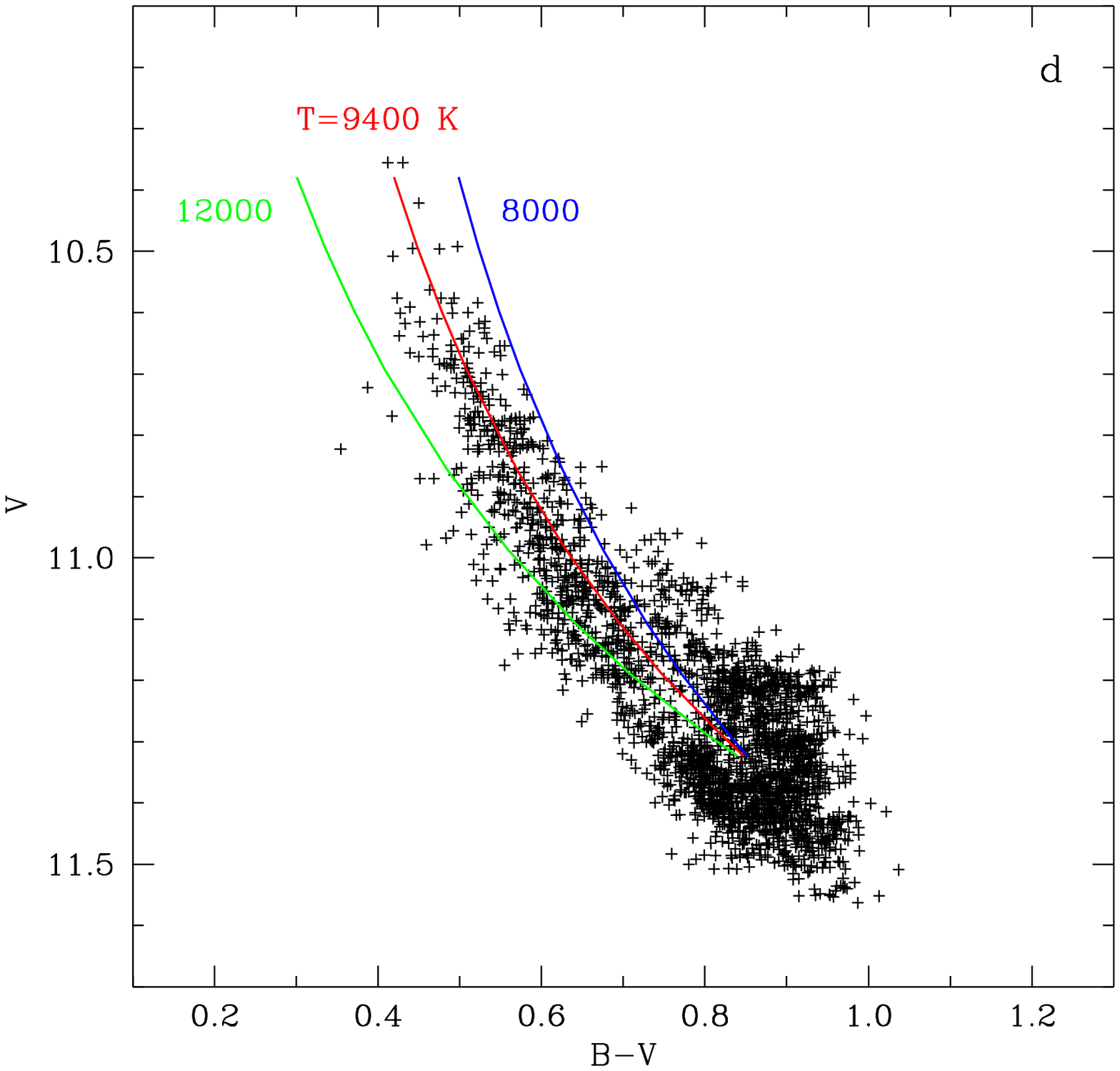}      
    \includegraphics{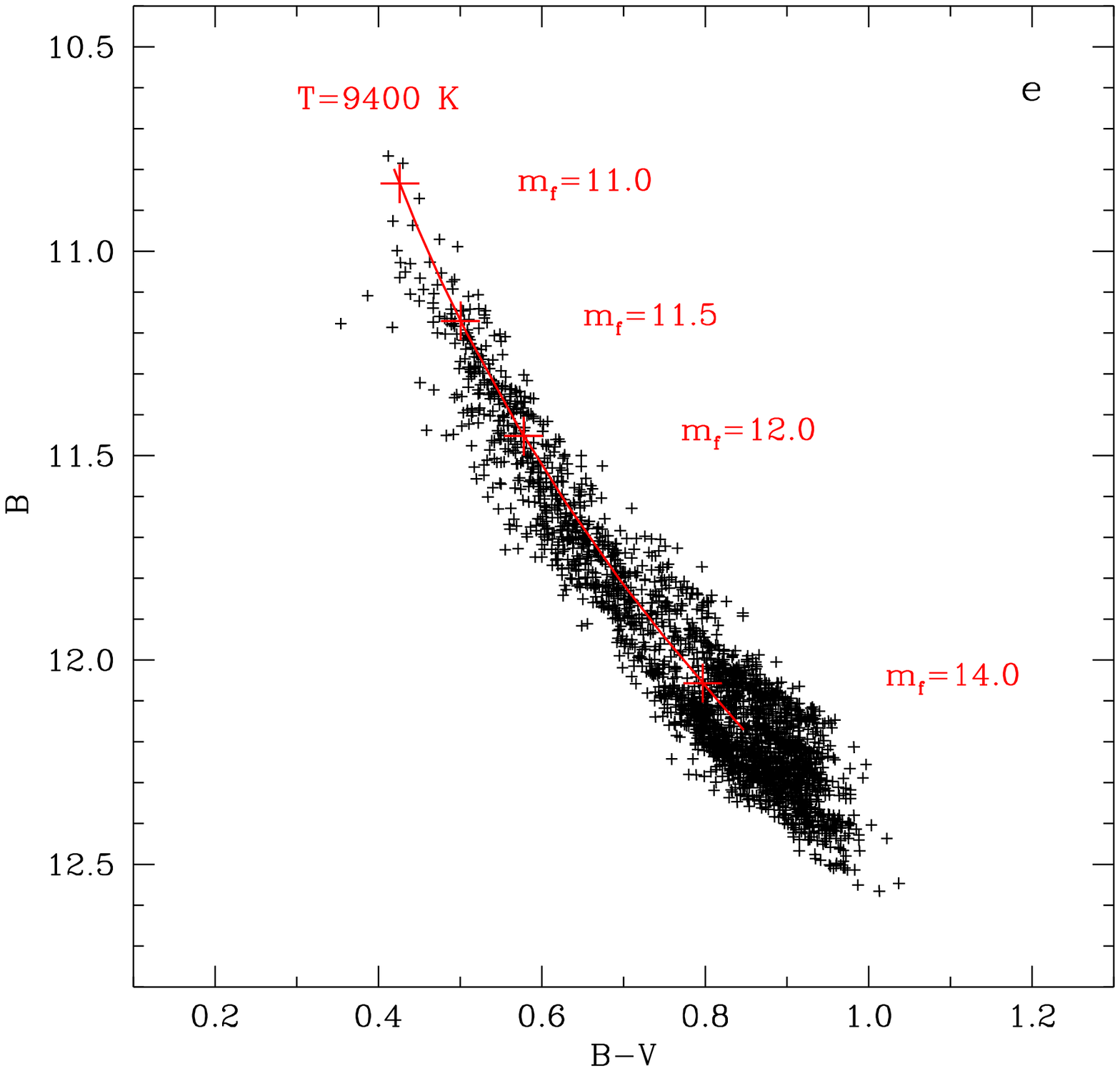}      
    \includegraphics{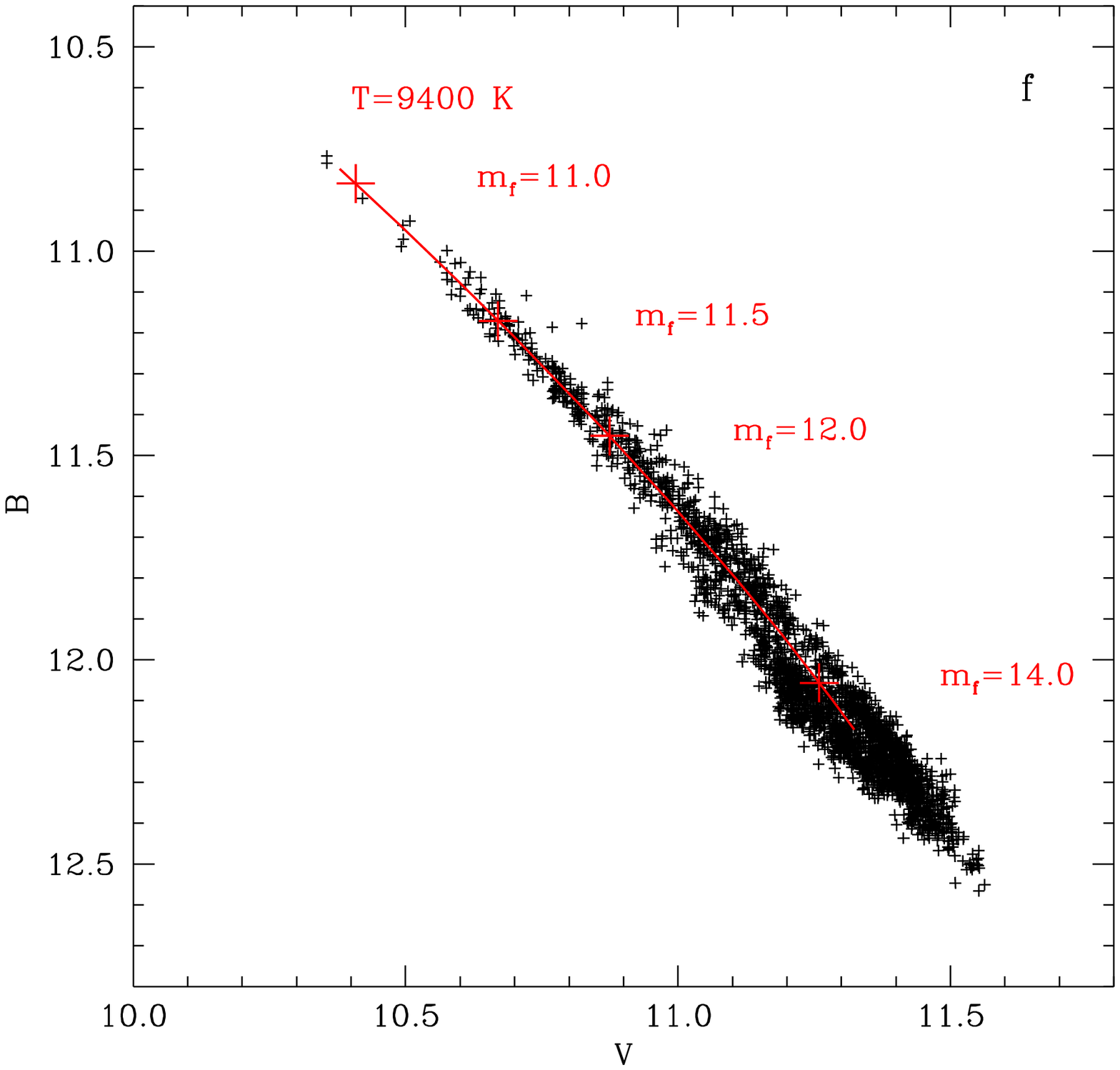}      
  \caption[]{Colour-magnitude diagrams  of the peculiar cataclysmic variable AE Aqr - 
    left panels: a) V versus  $B-V$, b) B versus $B-V$, c) B versus V.
    On the right panels the models are over-plotted. 
    }
  \label{fig.Cmag1}    
\end{figure*}	     

In Fig.~\ref{fig.Cmag1} are shown  the relation between the brightness of the star in B and V band 
and its $B-V$ colour. It is visible that  the star becomes bluer when it brightens and redder when it fades,
indicating that the flares change the colour of the system.


Pearson, Horne \& Skidmore (2003) have supposed that the flares of AE~Aqr are due to 
the ejection and expansion of  isothermal fireballs.
Using their improved model (Pearson, Horne \& Skidmore 2005),  
Zamanov et al. (2012) have  estimated 
the fireballs temperatures of 10000-25000 K, masses of $(7-90) \times 10^{19}$~g, 
and sizes of $(3-7) \times 10^9$ cm (using a distance of d=86 pc). 
These values refer to the peak of the flares observed in the UBVRI bands. 
Here we use E(B-V)=0 (La Dous 1991), and construct colour-magnitude diagrams. 
These diagrams should contain information about the temperature of the flares. 
We model the behaviour of AE~Aqr in the following way:   
\begin{itemize}
\item  The peak of the V band histogram  corresponds to V=11.35. 
       The peak of the B band histogram  corresponds to B=12.22. We adopt these values as basic level.
\item  To this level we add an additional source (fireball) with constant  $B-V$. 
       We vary the brightness of this additional source  from 15.0 mag to 10.95 magnitude in V. 
\item  Because the stellar magnitude scale is logarithmic,
during the calculations we convert the magnitudes into fluxes  using  Bessell (1979) calibration of a zero magnitude star. 
After it we reconvert the fluxes into magnitudes. 
\item  We repeat this procedure with different $B-V$ colour of the fireball.
We achieved the best agreement with  $B-V$=0.19 of the fireball.
\end{itemize}
The basic level is 
\begin{eqnarray}
  F_{V0}=3.610 \times  10^{-9}  \; \times \;  10^{-11.35/2.5}        \;  \;  \;  erg \: cm^{-2}  \: s^{-1} \:  \AA^{-1}  \\
  F_{B0}=6.601 \times  10^{-9}  \;  \times \; 10^{-12.22/2.5}        \;  \;  \;  erg \: cm^{-2}  \: s^{-1} \:  \AA^{-1},  
\end{eqnarray}
where   $F_{V0}$ and  $F_{B0}$ are the fluxes corresponding to the peak of the histograms in V and B band 
respectively.
The emission of the fireball is:  
\begin{eqnarray}
  F_{Vf}=3.610 \times  10^{-9}  \; \times \;  10^{-m_f/2.5}          \;  \;  \;  erg \: cm^{-2}  \: s^{-1} \:  \AA^{-1}  \\
  F_{Bf}=6.601 \times  10^{-9}  \;  \times \; 10^{-(m_f+c_f)/2.5}    \;  \;  \;  erg \: cm^{-2}  \: s^{-1} \:  \AA^{-1},  
\end{eqnarray}  
where $m_f$ is the V band magnitude of the fireball, $c_f$ is its B-V colour.
$F_{Vf}$ and  $ F_{Bf}$ are the fluxes of the ball emitted in V and B band respectively.
The brightness of the system ($F_V$ and  $F_B$)  is the sum of the fluxes of the basic level and the fireball:
\begin{eqnarray}
  F_V=F_{V0} +  F_{Vf} \\
  F_B=F_{B0} +  F_{Bf} \\
  m_V = -2.5 \; \log (F_V / (3.610 \times 10^{-9})) \\
  m_B = -2.5 \; \log (F_B / (6.601 \times 10^{-9})),
\end{eqnarray}
where   $ m_V$ and  $m_B$ are the  brightness of the system in magnitudes for V and B band respectively. 
In Fig.~\ref{fig.Cmag1}d we plot with green, red and blue colour three lines corresponding to B-V colour of the fireball 
$c_f =$ 0.03, 0.19 and  0.30. 
These B-V colours correspond to black body temperatures 12000~K, 9400~K and 8000~K,  respectively. 
The model with $c_f=0.19$  produces good agreement with the observations in the three panels:  
V versus  $B-V$  (Fig.~\ref{fig.Cmag1}d), B versus B-V (Fig.~\ref{fig.Cmag1}e) and 
B versus V (Fig.~\ref{fig.Cmag1}f). On  Fig.~\ref{fig.Cmag1}e and Fig.~\ref{fig.Cmag1}f with red crosses
are marked four positions corresponding to V brightness of the fireball 
$m_f= 11.0, \: 11.5, \: 12.0, \: 14.0$.

The largest  flares of AE~Aqr correspond to an additional source with  $V \approx 10.95$ mag, 
which is the brightness of the most luminous  flares (fireballs) in our data set. 
Comparing this value with the peak of the histogram  we derive that 
in V band the strongest flares emit 1.4 times as much 
energy as the non-flaring components of AE Aqr  (in the B band they emit 2.2 times). 
The values are similar but  higher than the value 1.5 times for the strongest flares observed in B band 
in 1983 (Bruch 1991).  
 

\section{Discussion}

AE Aqr is a highly variable object exhibiting  flaring behaviour  in the optical (Chincarini \& Walker 1981), 
radio (Bastian et al. 1988),  ultraviolet (Eracleous \& Horne 1996), and X-rays (Choi \& Dotani 2006). 
Five-colour (Walraven system) observations of the cataclysmic variable AE Aqr were made in 1984 and 1985 
(van Paradijs et al. 1989). 
The optical flux emitted in the flares is about 3 times the quiescent accretion flux. 
It seems to be a slight phase dependence of the activity -- while flares can definitely occur 
at any phase, the probability for very strong variations is higher in the first half of the orbital cycle 
than in the second half (Bruch \& Grutter 1997).

Different mechanisms are discussed to explain the flares of AE~Aqr:

(1) the flares are due to an accretion instability  that occurs within a few white-dwarf radii, perhaps as a result 
of magnetospheric gating of the inflowing matter at the inner edge of the accretion disc (van Paradijs et al. 1989). 

(2) blobs (fireballs) launched from the magnetosphere (Wynn et al. 1997; Pearson, Horne \& Skidmore 2002). 

In the context of the fragmented accretion flow/magnetic propeller scenario we assume  that the flares
represent the excitation of gaseous blobs upon encounter with the propeller, and their subsequent radiative 
cooling as they are expelled from the system. 
The gas stream emerging from L$_1$  encounters a rapidly spinning magnetosphere of the white dwarf. 
The rapid spin causes the stream material to be dragged toward  co-rotation with the magnetosphere. 
As it occurs outside  the corotation radius the magnetosphere accelerates velocity of the material beyond the escape velocity
(Wynn et al. 1997). 

The colour-magnitude diagrams (Fig.~\ref{fig.Cmag1}) indicate that  the balls have temperatures in the interval 
$8000   \lesssim T_f   \lesssim 12000$~K  and  V band  magnitude $14.0 \lesssim  m_f  \lesssim 11.0$.  
Our result point to that all the fireballs have similar temperature $\sim 9400$~K. 
Taking into account the observational errors and errors in the fit, we estimate the average temperature
of the fireballs  $9400\pm 600$~K. 
There is no tendency (at least  in our data) for the brighter flares to have higher or lower temperature. 
This is in agreement with the suppositions that the flares are isothermal (Pearson et al. 2003). 
This temperature is similar to that of the flickering source of the recurrent nova RS Oph
$T = 9500 \pm  500$~K (Zamanov et al. 2010).

\section{Conclusion}
We performed  37.5 hours simultaneous  observations in Johnson B and V bands of the cataclysmic variable 
AE~Aqr. On the basis of these observations
the $B-V$ colour is calculated for 2327 observational points.
The colour-magnitude diagrams show that the star becomes bluer as it becomes brighter. 
We modeled the brightness (B and V band) and  $B-V$  colour changes, supposing additional source of energy (blobs). 
The model indicates that the blobs (flares) have on average $B-V \approx 0.19$, which corresponds 
to an average temperature of the fireballs  $\approx 9400 \pm 600$~K. 

A related question is whether the colour-magnitudes diagrams and model discussed here 
are relevant to other cataclysmic and symbiotic stars with flickering.

\acknowledgements
This work is supported by the Program for career development of young scientists, 
Bulgarian Academy of Sciences. 


\begin{thebibliography}{}

\bibitem[Aleksi{\'c} et al.(2014)]{2014A&A...568A.109A} Aleksi{\'c}, J., Ansoldi, S., Antonelli, L.~A., et al.\ 2014, \aap, 568, A109 

\bibitem[Bastian et al.(1988)]{1988ApJ...324..431B} Bastian, T.~S., Dulk, G.~A., \& Chanmugam, G.\ 1988, \apj, 324, 431 

\bibitem[Belloni \& Stella(2014)]{2014SSRv..183...43B} Belloni, T.~M., \& Stella, L.\ 2014, \ssr, 183, 43 

\bibitem[Bessell(1979)]{1979PASP...91..589B} Bessell, M.~S.\ 1979, \pasp, 91, 589 

\bibitem[Bookbinder \& Lamb(1987)]{1987ApJ...323L.131B} Bookbinder, J.~A., \& Lamb, D.~Q.\ 1987, \apjl, 323, L131 

\bibitem[Bruch(1991)]{1991A&A...251...59B}   Bruch, A.\ 1991, \aap, 251, 59 

\bibitem[Bruch \& Grutter(1997)]{1997AcA....47..307B} Bruch, A., \& Grutter, M.\ 1997, \actaa, 47, 307 

\bibitem[Casares et al.(1996)]{1996MNRAS.282..182C} Casares, J., Mouchet, M., Martinez-Pais, I.~G., \& Harlaftis, E.~T.\ 1996, \mnras, 282, 182 

\bibitem[Chincarini \& Walker(1974)]{1974eaa..conf..249C} Chincarini, G., \& Walker, M.~F.\ 1974, 
In: Electrography and astronomical applications; Proceedings of the Conference  (A75-23926 09-89) 
Austin, University of Texas,  p. 249 

\bibitem[Chincarini \& Walker(1981)]{1981A&A...104...24C} Chincarini, G., \& Walker, M.~F.\ 1981, \aap, 104, 24 

\bibitem[Choi \& Dotani(2006)]{2006ApJ...646.1149C} Choi, C.-S., \& Dotani, T.\ 2006, \apj, 646, 1149 

\bibitem[Echevarr{\'{\i}}a et al.(2008)]{2008MNRAS.387.1563E} Echevarr{\'{\i}}a, J., Smith, R.~C., Costero, R., Zharikov, S., \& Michel, R.\ 2008, \mnras, 387, 1563 

\bibitem[Eracleous \& Horne(1996)]{1996ApJ...471..427E} Eracleous, M., \& Horne, K.\ 1996, \apj, 471, 427 

\bibitem[Henize(1949)]{1949AJ.....54...89H} Henize, K.~G.\ 1949, \aj, 54, 89 

\bibitem[Hill et al.(2016)]{2016MNRAS.459.1858H} Hill, C.~A., Watson, C.~A., Steeghs, D., Dhillon, V.~S., \& Shahbaz, T.\ 2016, \mnras, 459, 1858 

\bibitem[Ikhsanov et al.(2004)]{2004AstL...30..675I} Ikhsanov, N.~R., Neustroev, V.~V., \& Beskrovnaya, N.~G.\ 2004, Astronomy Letters, 30, 675 

\bibitem[Ikhsanov \& Beskrovnaya(2012)]{2012ARep...56..595I} Ikhsanov, N.~R., \& Beskrovnaya, N.~G.\ 2012, Astronomy Reports, 56, 595 

\bibitem[Isakova et al.(2016)]{2016ARep...60..498I} Isakova, P.~B., Ikhsanov, N.~R., Zhilkin, A.~G., Bisikalo, D.~V., \& Beskrovnaya, N.~G.\ 2016, Astronomy Reports, 60, 498 

\bibitem[Itoh et al.(2006)]{2006ApJ...639..397I} Itoh, K., Okada, S., Ishida, M., \& Kunieda, H.\ 2006, \apj, 639, 397 

\bibitem[Joy(1954)]{1954AJ.....59..326J} Joy, A.~H.\ 1954, \aj, 59, 326 

\bibitem[La Dous(1991)]{1991A&A...252..100L} La Dous, C.\ 1991, \aap, 252, 100 

\bibitem[Mauche et al.(2012)]{2012MmSAI..83..651M} Mauche, C.~W., Abada-Simon, M., Desmurs, J.-F., et al.\ 2012, \memsai, 83, 651 


\bibitem[Patterson et al.(1980)]{1980ApJ...240L.133P} Patterson, J., Branch, D., Chincarini, G., \& Robinson, E.~L.\ 1980, \apjl, 240, L133 

\bibitem[Pearson et al.(2002)]{2002ASPC..261..163P} Pearson, K.~J., Horne, K., \& Skidmore, W.\ 2002, The Physics of Cataclysmic Variables and Related Objects, 261, 163 

\bibitem[Pearson et al.(2003)]{2003MNRAS.338.1067P} Pearson, K.~J., Horne, K., \& Skidmore, W.\ 2003, \mnras, 338, 1067 

\bibitem[Pearson et al.(2005)]{2005ApJ...619..999P} Pearson, K.~J., Horne, K., \& Skidmore, W.\ 2005, \apj, 619, 999 

\bibitem[Scaringi(2015)]{2015AcPPP...2..107S} Scaringi, S.\ 2015, Acta Polytechnica CTU proceedings, Vol.~2,  107 

\bibitem[Schenker et al.(2002)]{2002MNRAS.337.1105S} Schenker, K., King, A.~R., Kolb, U., Wynn, G.~A., \& Zhang, Z.\ 2002, \mnras, 337, 1105 

\bibitem[Skidmore et al.(2003)]{2003MNRAS.338.1057S} Skidmore, W., O'Brien, K., Horne, K., et al.\ 2003, \mnras, 338, 1057 

\bibitem[Sokoloski(2003)]{2003ASPC..303..202S} Sokoloski, J.~L.\ 2003, Symbiotic Stars Probing Stellar Evolution, ASP Conf. 303, 202 

\bibitem[Torkelsson(2013)]{2013IAUS..290..145T} Torkelsson, U.\ 2013, Feeding Compact Objects: Accretion on All Scales,IAU Symp. 290, 145 

\bibitem[van Paradijs et al.(1989)]{1989A&AS...79..205V} van Paradijs, J., van Amerongen, S., \& Kraakman, H.\ 1989, \aaps, 79, 205 

\bibitem[Wynn et al.(1997)]{1997MNRAS.286..436W} Wynn, G.~A., King, A.~R., \& Horne, K.\ 1997, \mnras, 286, 436 

\bibitem[Zamanov et al.(2010)]{2010MNRAS.404..381Z} Zamanov, R.~K., Boeva, S., Bachev, R., et al.\ 2010, \mnras, 404, 381 

\bibitem[Zamanov  et al.(2012)]{2012AN....333..736Z} Zamanov , R.~K., Latev, G.~Y., Stoyanov, K.~A., et al.\ 2012, Astronomische Nachrichten, 333, 736 

\end{thebibliography}
\end{document}